\documentclass[twocolumn,aps,prl]{revtex4-1}

\usepackage{epsfig, color}
\usepackage{amsthm, amsmath, amsfonts, ae, psfrag}
\usepackage{braket}
\usepackage{bbm}
\usepackage[english]{babel}
\usepackage{epstopdf} 
\usepackage{listings}
\usepackage{placeins}	
\usepackage{verbatim} 
\usepackage{framed}
\usepackage{xcolor} 
\colorlet{shadecolor}{gray!25} 

\usepackage[	format			= plain,
				font			= footnotesize, 
				justification	= centerlast	]{caption}
\usepackage[]{subcaption}

\usepackage{babel}
\usepackage{changes}

\begin{document}

\title{Multivalent Ion-Activated Protein Adsorption Reflecting Bulk Reentrant Behavior}

\author{Madeleine R. Fries\textsuperscript{1},
\ Daniel Stopper\textsuperscript{2}, 
\ Michal K. Braun\textsuperscript{1}, 
\ Alexander Hinderhofer\textsuperscript{1},
\ Fajun Zhang\textsuperscript{1},
\ Robert M. J. Jacobs\textsuperscript{3},
\ Maximilian W. A. Skoda\textsuperscript{4},
\ Hendrik Hansen-Goos\textsuperscript{2},
\ Roland Roth\textsuperscript{2}
\ and Frank Schreiber\textsuperscript{1}}

\affiliation{\textsuperscript{1} Institute for Applied Physics, University of T{\"u}bingen, 72076 T{\"u}bingen, Germany}
	\affiliation{\textsuperscript{2} Institute for Theoretical Physics, University of T{\"u}bingen, 72076 T{\"u}bingen, Germany}
	
	\affiliation{\textsuperscript{3} Department for Chemistry, Chemistry Research Laboratory, University of Oxford, Oxford, OX1 3TA, United Kingdom}
	
	\affiliation{\textsuperscript{4} Rutherford-Appleton Laboratory, ISIS Facility, Didcot, OX11 0QX, United Kingdom}

\date{\today}

\begin{abstract}
Protein adsorption at the solid-liquid interface is an important
phenomenon that often can be observed as a first step in biological processes. 
Despite its inherent importance, still relatively little is known about the underlying microscopic mechanisms. 
Here, 
using multivalent ions,
we demonstrate the control of the interactions and the corresponding adsorption of net-negatively charged proteins (bovine serum albumin) 
at a solid-liquid interface. This is demonstrated by ellipsometry and corroborated by neutron reflectivity and quartz-crystal microbalance experiments.
We show that the reentrant condensation observed within the rich {\it bulk} phase behavior of the system featuring a nonmonotonic dependence of the second virial coefficient on salt concentration $c_s$ 
is reflected in an intriguing way in the protein adsorption $d(c_s)$ at the {\it interface}. 
Our findings are successfully described and understood by a model of ion-activated patchy interactions 
within the framework of classical density functional theory. 
In addition to the general challenge of connecting bulk and interface behavior, our work has implications
for, {\it inter alia}, nucleation at interfaces. 
\end{abstract}


\maketitle

The interactions of proteins, with their inherent heterogeneity and differently charged patches, 
in addition to hydrophilic and hydrophobic regions and dispersion forces, 
are very complex \cite {Rabe11}. 
While obviously required for their biological function,
this complexity of the interactions is very demanding for a quantitative physical understanding.
Particularly difficult is the connection to the associated mesoscopic and macroscopic behavior,
with enormous implications for a range of rather diverse fields. These include the understanding of protein crystallization \cite {Wolde97, Whitelam2012, Fusco2013}, 
as well as various forms of aggregation \cite{stradner04,godfrin2015},
whether biologically desired 
\cite{schmidt08,yan13,monks98}
or related to diseases such as Alzheimer's
\cite{kakio01}, Huntington's, or prion diseases (e.g. Creutzfeldt-Jakob disease) \cite{aguzzi10}. 
%
%
A further level of complexity is added by the frequently heterogeneous environment in soft and biological systems, 
often with internal interfaces, at which adsorption might take place, coexisting with fluid (bulk-like) regions.
While in numerous studies the phase behavior of proteins has been investigated \cite{stradner04, Sear1999, Goegelein2008}, 
it remains an important challenge to understand protein-protein interactions in a microscopic picture 
and to predict the resulting macroscopic thermodynamic behavior of proteins in solution and at interfaces, and how these behaviors correspond or differ.

\begin{figure} 
\includegraphics[width=0.5\textwidth]{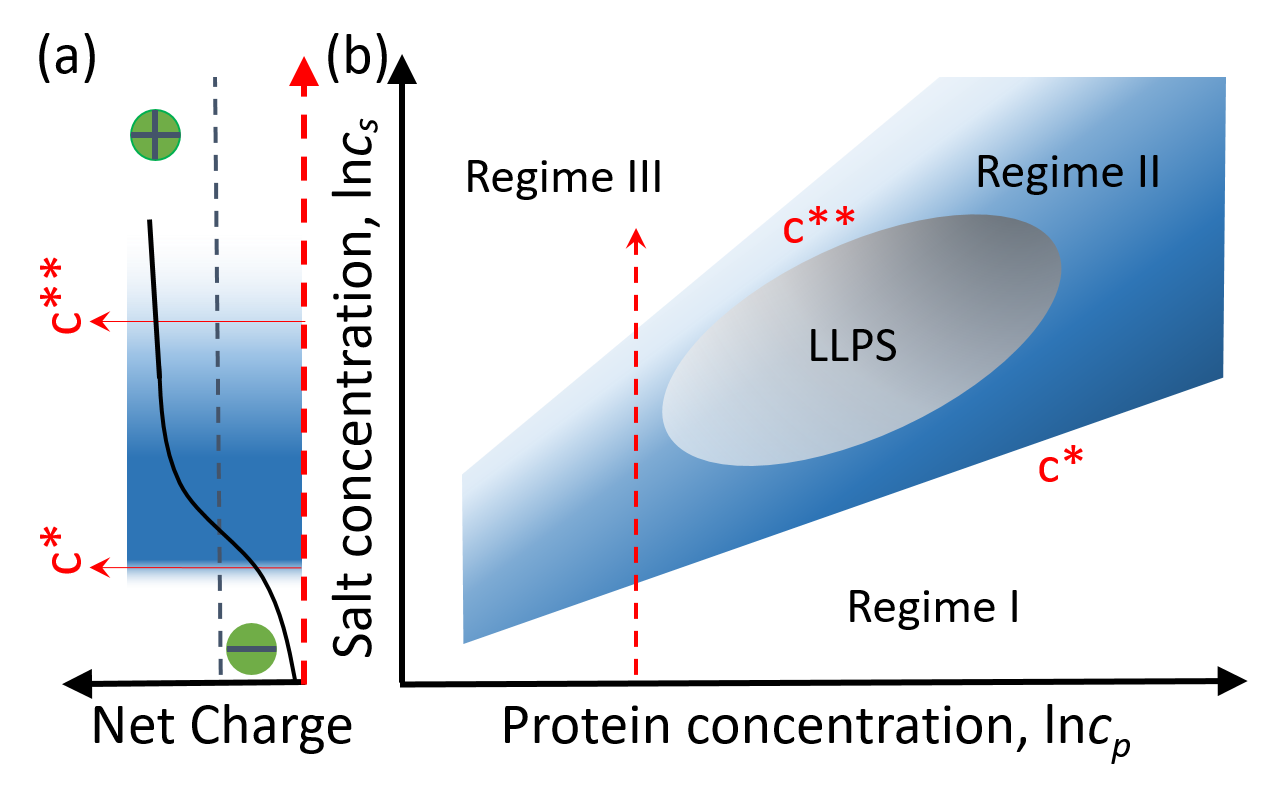}
\caption{(a) Charge inversion of BSA as a consequence of adding
  trivalent yttrium ions. (b) Schematic of the bulk phase diagram of
  BSA and YCl$_3$ (modified from Ref. \cite{Zhang14}) showing a
  liquid-liquid phase separation (LLPS) and reentrant
  condensation. The dashed red arrow indicates the path taken in the
  experiments.}
\label{fig:phase_diagram}
\end{figure}

For the manipulation of the bulk phase behavior different strategies have been demonstrated.
On the one hand, the use of co-solvents such as glycerol to stabilize a protein solution \cite{cosolvent} can
help to avoid protein aggregation and cluster formation. On the other hand, enzymatic crosslinking  \cite{saricay12} or the use of trivalent ions such as yttrium cations can be employed to trigger bridge formation between globular proteins, which can lead to cluster formation, reentrant condensation and liquid-liquid phase separation (Fig. \ref{fig:phase_diagram}) \cite{Zhang08, Zhang12}.

Protein adsorption at solid-liquid interfaces occurs in many natural processes, and its understanding is crucial in many fields, 
ranging from biotechnology, biology, pharmacology, and medicine to environmental science and food processing with relevance in many applications 
\cite{Rabe11}. 
In particular, it is the first step in numerous biological processes, such as the blood coagulation cascade,
transmembrane signaling and adhesion of particles (bacteria or cells)
\cite {Rabe11}, and therefore plays a key role in biomedical
devices, including biosensors, biochips, soft contact lenses and
biomaterials for implants \cite{Castner17}.

Bovine serum albumin (BSA) is considered as one of the model proteins for adsorption studies \cite{schottler16}.
In solution, BSA is a globular protein with well-characterized physico-chemical properties \cite{Su99}. 
Serum albumin is the most abundant blood protein in mammals, and its adsorption has been intensely studied with different methods,
under various conditions \cite{Rabe11, Lu07, Gray04, SM}.
Nevertheless, {\it controlling} the interactions and connecting to the bulk behavior remains a challenge. 
In that context, the use of multivalent ions  \cite{Kandori10, Anbazhagan16, LeeSchmidtFenter2013} offers a viable path, 
with the unique opportunity to tailor and even invert the charge state of proteins as well as surfaces by overcompensation \cite{Zhang08, Zhang12, Sauter15}, which has been demonstrated to be a rather universal approach \cite{zhang10}.


In this Letter, we demonstrate the use of multivalent ions (Y$^{3+}$) to control the interaction of BSA with SiO$_2$ interfaces. 
We find reentrant {\it interface} adsorption behavior, reflecting in an intriguing way the {\it bulk} phase behavior [Fig. \ref{fig:phase_diagram} (b)].
Furthermore, we show that both bulk and interface adsorption behavior can be modeled
consistently by statistical mechanics of ion-activated patches \cite{RoosenRunge2014}. 

\begin{figure*}
\includegraphics[width=0.9\textwidth]{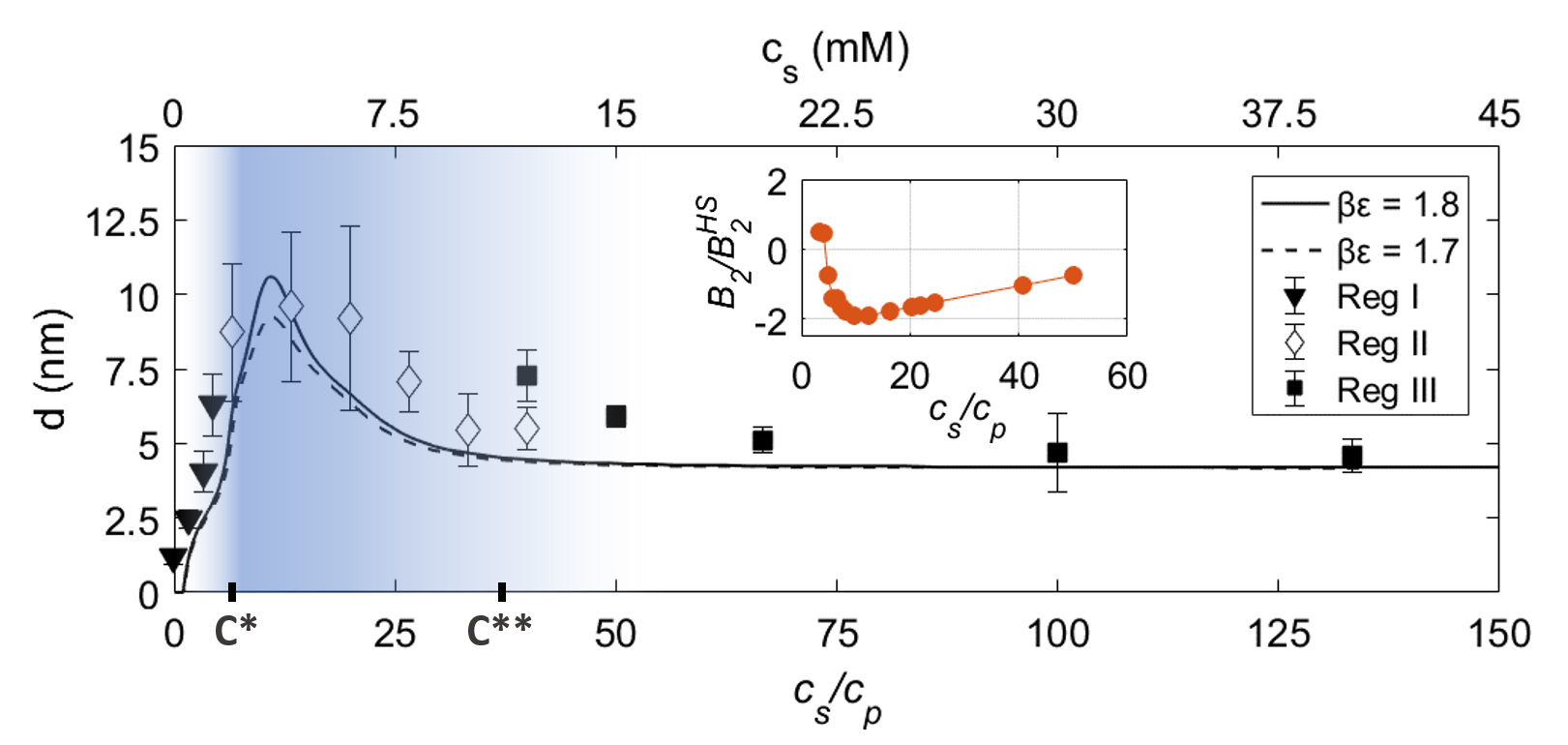}
\caption{$Individual$ $symbols$: Adsorbed protein layer thickness $d$ extracted from ellipsometry
  as a function of $c_s/c_p$. $c^*$ and $c^{**}$ denote the phase transitions of the bulk solution \cite{Zhang08} [Fig. \ref{fig:phase_diagram} (b)]. Note that around $c^{**}$, there is an experimental difference between the data in regime III vs. regime
  II. The centrifuged samples in regime II reflect the adsorption trend
  for overall lower adsorption values due to the removal of big clusters in bulk solution,
  but still follow the same adsorption trend. In addition, the top $c_s$-axis is included showing the absolute $c_s$ in the system (at $c_p$ = 20 mg/mL). The blue shaded area shows the approximate range of the bulk turbidity. $Solid$ $and$ $dashed$ $lines$: Protein adsorption based on DFT calculations as born out by the ion-activated attractive patch model, while neglecting long-range forces, as a function of $c_s/c_p$
  for two different values of $\beta\varepsilon$. $Inset$: $B_2/B_2^H{}\!^S$ is the reduced second virial coefficient obtained $via$ SAXS measurements. 
}
\label{fig:Adsorption_trend}
\end{figure*}

BSA (molecular weight $M_W$ = 66 kDa) and YCl$_3$ were obtained from Sigma Aldrich and used as received. BSA is net negatively charged above its isoelectric point of pH = $4.6$ \cite {Zhang08}. Protein solutions were prepared by mixing the stock solutions at temperature $T$ = 20 $^{\circ}$C. The working protein concentration $c_p$ was set to 20 mg/mL and the trivalent salt concentration $c_s$ ranged from 0.5 - 40 mM [depicted by the red arrow in Fig. \ref{fig:phase_diagram} (b)]. With increasing $c_s$, protein solutions undergo a reentrant condensation (RC) phase behavior in regime III. An aggregation regime II occurs in between two salt concentrations, $c^*$ and $c^{**}$ as illustrated in Fig. \ref{fig:phase_diagram} (a-b). The physical mechanisms behind the observed RC behavior are the effective inversion of protein charge [Fig. \ref{fig:phase_diagram} (a)] and a cation-mediated anisotropic attraction \cite{RoosenRunge2014,Zhang08}. The effective interactions $V_{ \mathrm{eff}(r)}$ between proteins are reflected in the behavior of the reduced second virial coefficient $B_2/B_2^H{}\!^S$. $B_2$ defines the second viral coefficient of the $bulk$ solution
\begin{equation}
B_2 = 2 \pi \int_0^\infty \text{d}r\,r^2\left[ 1-e^{-\beta \,V_\text{eff} (r)} \right].
\end{equation}
The second viral coefficient of hard spheres is defined by $B_2^H{}\!^S$ = $16 \pi R_p^3/3$, where $R_p$ is the radius of the protein. Experimental $B_2/B_2^H{}\!^S$ (orange inset, Fig. \ref{fig:Adsorption_trend}) were determined using small-angle X-ray scattering (SAXS) (ID02 at the ESRF in Grenoble, France) \cite{Braun17}.

The adsorption studies were performed on standard Si wafers with native oxide
layer. Before each measurement, all components of the liquid cell were
cleaned at 50 $^{\circ}$C $via$ ultrasonication in acetone,
isopropanol, and degassed water for 10 min in each
solvent. Ellipsometry (Woollam VASE M-2000 and
Beaglehole Picometer) was
employed {\it in situ} at the Brewster angle of 68$^{\circ}$ (for
SiO$_2$) to extract an effective protein layer thickness $d$, assuming a Cauchy layer with density corresponding to that of pure BSA (i.e. volume fraction of 1; see Supporting Material \cite{SM} for definition of $d$, which includes Refs. \cite{Hlady88, Tilton90, Rabe09, Su98, Rabe08, Larsericsdotter05, Giacomelli01, Tarasevich02, Roach05, mondon03, Elwing98, Tsargorodskaya04, Sauter11, herzinger98, schiebener90, voinova02, feiler07}). Complementary studies were performed using neutron
reflectometry (NR) at the INTER beam line at ISIS (Rutherford Laboratory,
Didcot, UK) \cite{Webster06, Skoda09}, as well as quartz crystal
microbalance (QCM, Q-Sense Analyzer Biolin Scientific), confirming
the trends in the adsorption behavior $d(c_s)$ \cite{SM}.  For better comparability, the thicknesses extracted from NR and QCM-D are also normalized to an assumed BSA volume fraction of 1 \cite{SM}.

Based on real-time ellipsometric data of the adsorption kinetics, we
extract $d$ in the long-time limit (saturation after $\sim 60$ min) and plot it in Fig.~\ref{fig:Adsorption_trend} as a function of  $c_s/c_p$. It is convenient to use a dimensionless salt axis, i.e. {$c_s/c_p$}, especially when comparing to theory. Both BSA and SiO$_2$ surfaces are net negatively charged in water (no added salt). Under these conditions the electrostatic repulsion among the proteins dominates the solution compared to the repulsion between the proteins and the solid surface leading to a minimum of the protein adsorption. Evaluation of the ellipsometric data shows that then only a $d$ of $1.2 \pm 0.25$ nm is adsorbed. Upon increasing $c_s$ to 1.3 mM still in the clear regime I as illustrated in Fig. \ref{fig:phase_diagram} (b), $d$ increases to $6.29 \pm 1.02$ nm (solid triangles in Fig.~\ref{fig:Adsorption_trend}). In our system, we assume a $R_p$ of $\sim 3.5$ nm \cite{Yohannes10,SM} and define one monolayer equivalent (ML) to be $d \approx 4$ nm \cite{SM}, corresponding in regime I at 1.3 mM to the formation of $d$ > 1 ML. 
 
In regime II, $d$ increases towards a maximum value of
$9.59 \pm 2.5$ nm (> 1 ML) at $c_s$ = 4 mM (empty diamonds, Fig. \ref{fig:Adsorption_trend}). At still higher $c_s$,
$d$ decreases down to $\sim 6$ nm approaching the
upper boundary of regime II at $c^{**}$. Note that in regime II (empty diamonds) the
bulk solution is centrifuged before the adsorption experiments, which
explains the jump of $d$ in the transition region between
regime II and III. This is done because the solution in regime II is too turbid due to extensive protein cluster formation, which causes massive bulk light scattering and a lack of sensitivity of the ellipsometer. We show both data sets at $c_s/c_p$ = 40 (centrifuged and non-centrifuged) to account for the experimental difference, which, importantly, does not affect the overall adsorption trend. 

In regime III close to $c^{**}$, $d$ is $7.28 \pm 0.87$ nm at $c_s$ = 12 mM, but with increasing $c_s$, $d$ decreases down to a plateau value of 4.5 nm above 30 mM (solid squares in Fig.~\ref{fig:Adsorption_trend}). $d$ then corresponds to slightly less than one full ML. These experimental results are supported by complementary measurements (NR and QCM) \cite {SM}. It is interesting to note that after rinsing with pure water the surface retains an irreversibly bound layer of protein with $d$ = 4 nm.

To understand the adsorption behavior, it is important to
realize that the behavior of $d$ is closely related to that of $B_2/B_2^H{}\!^S$ of the {\it bulk} solution (inset, Fig. \ref{fig:Adsorption_trend}). In regime II, the value of $B_2/B_2^H{}\!^S$ is clearly negative indicating a strong overall attraction between proteins compared to regime I and III. Note that this is {\it not} the definition of the regimes nor its boundaries, but rather is an important observation. The net attraction between proteins is reflected by a sharp adsorption maximum.
This observation indicates that the protein adsorption in our system is closely related to the bulk behavior, which can successfully be accounted for
by the model for ion-activated attractive patches as a mechanism for 
interactions in protein-salt mixtures \cite{RoosenRunge2014}. This model is formulated within the Wertheim theory for associating fluids 
\cite{Jackson1988, Chapman1988,WertheimI1984, WertheimII1984, WertheimIII1986,WertheimIV1986,romano2010phase,russo2011re,pawar2010fabrication}, and treats proteins as hard spheres with radius $R_p$ and $M$ distinct 
and independent binding sites (patches) \cite{pawar2010fabrication}. These 
sites can be occupied by salt ions, thereby activating a given patch (ion binding). The 
occupation probability of a site is given by  
$\Theta = (1 + \exp(\beta\varepsilon_b - \beta\mu_s))^{-1}$, where 
$\mu_s$ denotes the salt chemical potential, $\beta= (k_B T)^{-1}$, and 
$\varepsilon_b$ the binding energy \cite{RoosenRunge2014}. A bond between 
two patches of distinct proteins is possible only if an activated patch 
meets a de-activated one (ion bridge). As a result, $c_s$ controls 
the protein-protein interactions. Note, however, that only the proteins 
are represented explicitly in this model. This implies that $c_s$ as a function of $\mu_s$ cannot be predicted self-consistently within this approach. We use the
location of the minimum of the experimentally determined $B_2/B_2^H{}\!^S$ in order to calibrate $c_s(\mu_s)$. 

The resulting phase diagram of the model accounts for key features of the rather rich 
experimental phase diagram, such as reentrant condensation and a 
closed-loop LLPS binodal schematically shown in Fig.~\ref{fig:phase_diagram} (b) \cite{RoosenRunge2014}.   The model also allows predictions of regions in the phase diagram which are populated by protein
clusters. A quantitative measure for this is $\Phi$, the fraction of
proteins in clusters. In the present study we assume that in region II at
least $20\%$ of the proteins are part of
clusters, i.e., $\Phi=0.2$ to define $c^*$ and $c^{**}$.  

While the experimental results presented here suggest that the bulk
behavior dominates the adsorption trend, the  
key point in the present study is the protein adsorption at a
charged planar wall, which implies breaking the translational symmetry of the system. To this end, we employ classical density functional theory
(DFT) \cite{Evans1979} which provides a powerful and well-established
framework to investigate inhomogeneous density distributions. Within
DFT one can show rigorously \cite{Evans1979} that a functional
\begin{equation}
\Omega[\rho] = {\cal F}[\rho] + \int \text{d}\mathbf{r}\,\rho({\bf r}) \left[
V_\text{ext}({\bf r}) - \mu \right]
\end{equation}
of the inhomogeneous density profile $\rho({\bf r})$ exists and takes
its minimum, the grand potential, at the equilibrium density 
distribution.

\begin{figure}[t!] 
  \begin{subfigure}{0.5\textwidth}
    \includegraphics[width=8.5cm]{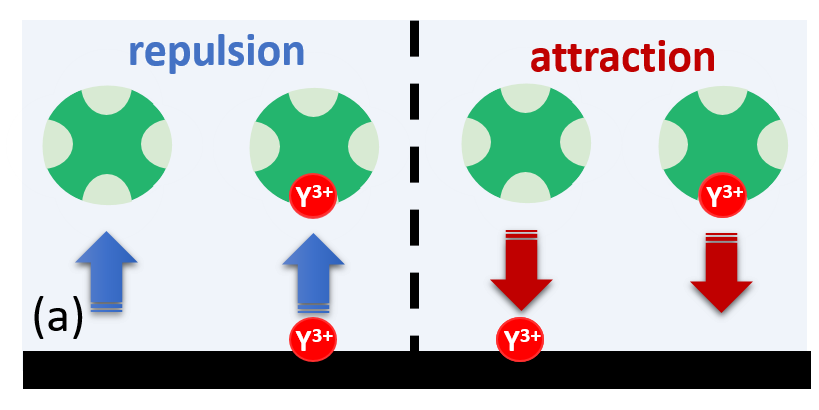}
    \label{binding_mechanism}
  \end{subfigure}
  \begin{subfigure}{0.5\textwidth}
    \includegraphics[width=9cm]{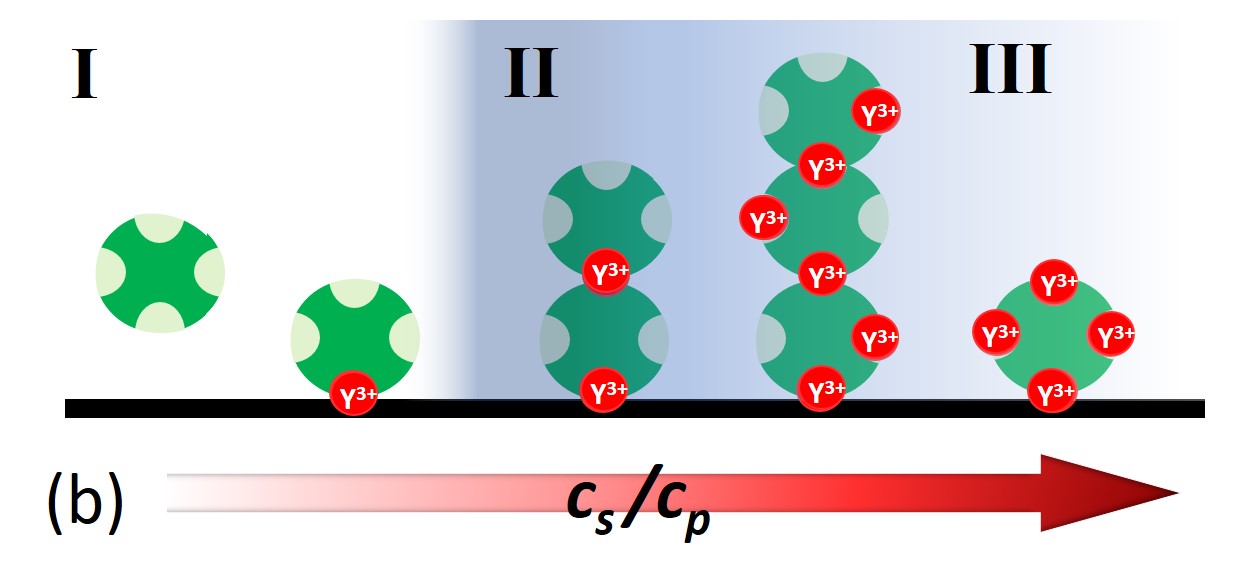}
   
    \label{sketch}
  \end{subfigure}
  \caption{(a) Illustration of the different interaction mechanisms of
    the proteins, salt, and interface. (b) Sketch of protein adsorption on the attractive surface by increasing $c_s/c_p$.}
  \label{fig:sketch}
\end{figure}

Using a DFT formulation of the Wertheim theory \cite{YuWu2002} based 
on fundamental measure theory (FMT) for hard spheres \cite{Rosenfeld1989,
  Roth2010}, we calculate $d$ at the SiO$_2$-water 
interface. This interface is charged and strongly attracts yttrium ions, 
which in turn attract proteins towards the wall [Fig.~\ref{fig:sketch} (a)]. Effectively, this can be described
by a short-ranged external potential $V_\text{ext}(z)$ acting on the 
proteins, where $z$ is the distance normal to the SiO$_2$ wall. We set
$\beta V_\text{ext}(z) = \infty$ for $z < 0$ in order to represent a
steric repulsion between proteins and the substrate and $\beta
V_\text{ext}(z) = -\beta\varepsilon M \Theta \xi(z)$ for $z \geq
0$. $\xi(z)$ accounts for the rather short-ranged attraction
induced by the yttrium ions {\em condensed} on the wall -- which is in-line with recent experimental observations
\cite{LeeSchmidtFenter2013}. Here, we employ a Gaussian form $\xi(z) =
\exp(-0.5 (z/R_p)^2)$ with the range of attraction being roughly one
protein diameter, which {\em effectively} accounts for the range
of the screened electrostatic interactions between ions and the
wall, and between ions and proteins.

The strength of the external potential depends on $\mu_s$ $via$ the occupation probability $\Theta$ of
the protein binding sites. This form can be motivated by the
following arguments. A sketch is presented in Fig.~\ref{fig:sketch}. At 
low $c_s$, when $\Theta \to 0$, only a few proteins are subjected
to the attraction of the wall induced by the ions. As $c_s$ increases, more
ions mediate the attractions between the wall and the proteins. At the
same time the protein-protein attraction increases accordingly, which
leads in turn to an increase in $d$. At very high $c_s$ ($\Theta \to 1$), 
the mechanism for the wall attraction remains, while the
protein-protein interaction becomes weak since a majority of the binding sites are occupied so that salt ions can no longer cause a patchy attraction between the proteins. Therefore, one expects from our 
model $\sim 1$ ML of proteins to be adsorbed on the wall for $\Theta \to 1$.

In Fig.~\ref{fig:Adsorption_trend} (solid and dashed lines), we show the value of $d$ in nm as a function of $c_s/c_p$ for a volume packing fraction 
$\eta = (4 \pi /3) \rho_p R_p^3 = 0.0078$, corresponding
to $c_p=20$ mg/mL along the path indicated by the red dashed arrow in
Fig.~\ref{fig:phase_diagram} (b). We choose  $M = 4$ and $\varepsilon_b = -5$ ~\cite{RoosenRunge2014}. The protein adsorption is
computed from the inhomogeneous density profile $\rho_p(z)$, obtained $via$ the DFT for our activated patch model. In order to compare
to experiments, we define $d$ as the distance from the wall where $\rho_p(z)$ is at least 50$\%$
higher than the bulk density $\rho_p$. For suitable values of 
$\beta\varepsilon$ [$1.8$ (solid curve), and $1.7$ (dashed curve)], we
find very good, semi-quantitative agreement between theory and 
experiment. For high values of $c_s$, we find a finite $d$ related to $\sim 1$ML similar to the experiments. Note that the fraction $\Phi$ of proteins in clusters in the bulk system is directly related to the behavior of the layer thickness $d$ of proteins at the wall.
 
Our theoretical results confirm that ion binding at the protein
surface drives the experimentally observed non-monotonic adsorption
behavior, thereby reflecting the underlying bulk interactions. In
particular, the remarkable agreement between experiment and theory
(considering in particular the few parameters involved) emphasizes that our model of
ion-activated attractive patchy particles, subjected to an effective external
wall potential, captures the essential effects of the protein
adsorption at a charged surface in the presence of multivalent salt ions. 
Our model is kept intentionally simple with a minimum number of
parameters, which helps us to identify the key mechanism responsible
for the behavior of the system, namely the ion-activated patchy
interactions of the proteins. Importantly, using our model we can explore the
adsorption behavior of our system in different parts of the bulk phase
diagram. As we increase the protein concentration approaching the
LLPS region the adsorbed film thickness $d$ increases. We find a 
{\em complete} wetting regime in which $d$ becomes even macroscopically
thick \cite {SM}. {\em Qualitatively} we find similar behavior as
shown in Fig.~\ref{fig:Adsorption_trend} with a maximum for $c^*$<$c_s$<$c^{**}$.

In conclusion, we have demonstrated that multivalent ions can be
employed not only to control the bulk interactions and bulk phase 
behavior of proteins such as BSA, but also its adsorption behavior
at a charged interface such as water-SiO$_2$. We observe  
reentrant effects at the interface, which reflects the bulk behavior,
measured by $B_2/B_2^H{}\!^S$, in an intriguing way. 
Furthermore, the experimental data can be explained and understood 
by theoretical calculations within the framework of classical DFT 
based on a model of ion-activated patchy interactions and their
associated statistics. In addition to the fundamental implications of
the first-time demonstration of this ion-activated patch model in the
context of the symmetry break brought about by an interface, our approach may pave the way
to controlled nucleation at interfaces in regime II and possibly
protein crystallization under new conditions. 

\begin{acknowledgments}
Funding by the DFG and Carl-Zeiss-Stiftung is gratefully acknowledged. 
\end{acknowledgments}

\bibliographystyle{apsrev4-1}
\bibliography{letterProteinAdsorption}

\begin{thebibliography}{64}%
\makeatletter
\providecommand \@ifxundefined [1]{%
 \@ifx{#1\undefined}
}%
\providecommand \@ifnum [1]{%
 \ifnum #1\expandafter \@firstoftwo
 \else \expandafter \@secondoftwo
 \fi
}%
\providecommand \@ifx [1]{%
 \ifx #1\expandafter \@firstoftwo
 \else \expandafter \@secondoftwo
 \fi
}%
\providecommand \natexlab [1]{#1}%
\providecommand \enquote  [1]{``#1''}%
\providecommand \bibnamefont  [1]{#1}%
\providecommand \bibfnamefont [1]{#1}%
\providecommand \citenamefont [1]{#1}%
\providecommand \href@noop [0]{\@secondoftwo}%
\providecommand \href [0]{\begingroup \@sanitize@url \@href}%
\providecommand \@href[1]{\@@startlink{#1}\@@href}%
\providecommand \@@href[1]{\endgroup#1\@@endlink}%
\providecommand \@sanitize@url [0]{\catcode `\\12\catcode `\$12\catcode
  `\&12\catcode `\#12\catcode `\^12\catcode `\_12\catcode `\%12\relax}%
\providecommand \@@startlink[1]{}%
\providecommand \@@endlink[0]{}%
\providecommand \url  [0]{\begingroup\@sanitize@url \@url }%
\providecommand \@url [1]{\endgroup\@href {#1}{\urlprefix }}%
\providecommand \urlprefix  [0]{URL }%
\providecommand \Eprint [0]{\href }%
\providecommand \doibase [0]{http://dx.doi.org/}%
\providecommand \selectlanguage [0]{\@gobble}%
\providecommand \bibinfo  [0]{\@secondoftwo}%
\providecommand \bibfield  [0]{\@secondoftwo}%
\providecommand \translation [1]{[#1]}%
\providecommand \BibitemOpen [0]{}%
\providecommand \bibitemStop [0]{}%
\providecommand \bibitemNoStop [0]{.\EOS\space}%
\providecommand \EOS [0]{\spacefactor3000\relax}%
\providecommand \BibitemShut  [1]{\csname bibitem#1\endcsname}%
\let\auto@bib@innerbib\@empty
\bibitem [{\citenamefont {Rabe}\ \emph {et~al.}(2011)\citenamefont {Rabe},
  \citenamefont {Verdes},\ and\ \citenamefont {Seeger}}]{Rabe11}%
  \BibitemOpen
  \bibfield  {author} {\bibinfo {author} {\bibfnamefont {M.}~\bibnamefont
  {Rabe}}, \bibinfo {author} {\bibfnamefont {D.}~\bibnamefont {Verdes}}, \ and\
  \bibinfo {author} {\bibfnamefont {S.}~\bibnamefont {Seeger}},\ }\href@noop {}
  {\bibfield  {journal} {\bibinfo  {journal} {Adv. Colloid Interface Sci}\
  }\textbf {\bibinfo {volume} {162}},\ \bibinfo {pages} {87} (\bibinfo {year}
  {2011})}\BibitemShut {NoStop}%
\bibitem [{\citenamefont {Rein~ten Wolde}\ and\ \citenamefont
  {Frenkel}(1997)}]{Wolde97}%
  \BibitemOpen
  \bibfield  {author} {\bibinfo {author} {\bibfnamefont {P.}~\bibnamefont
  {Rein~ten Wolde}}\ and\ \bibinfo {author} {\bibfnamefont {D.}~\bibnamefont
  {Frenkel}},\ }\href@noop {} {\bibfield  {journal} {\bibinfo  {journal}
  {Science}\ }\textbf {\bibinfo {volume} {277}},\ \bibinfo {pages} {1975}
  (\bibinfo {year} {1997})}\BibitemShut {NoStop}%
\bibitem [{\citenamefont {Whitelam}(2010)}]{Whitelam2012}%
  \BibitemOpen
  \bibfield  {author} {\bibinfo {author} {\bibfnamefont {S.}~\bibnamefont
  {Whitelam}},\ }\href@noop {} {\bibfield  {journal} {\bibinfo  {journal}
  {Phys. Rev. Lett.}\ }\textbf {\bibinfo {volume} {105}},\ \bibinfo {pages}
  {088102} (\bibinfo {year} {2010})}\BibitemShut {NoStop}%
\bibitem [{\citenamefont {Fusco}\ and\ \citenamefont
  {Charbonneau}(2013)}]{Fusco2013}%
  \BibitemOpen
  \bibfield  {author} {\bibinfo {author} {\bibfnamefont {D.}~\bibnamefont
  {Fusco}}\ and\ \bibinfo {author} {\bibfnamefont {P.}~\bibnamefont
  {Charbonneau}},\ }\href@noop {} {\bibfield  {journal} {\bibinfo  {journal}
  {Phys. Rev. E}\ }\textbf {\bibinfo {volume} {88}},\ \bibinfo {pages} {012721}
  (\bibinfo {year} {2013})}\BibitemShut {NoStop}%
\bibitem [{\citenamefont {Stradner}\ \emph {et~al.}(2004)\citenamefont
  {Stradner}, \citenamefont {Sedgwick}, \citenamefont {Cardinaux},
  \citenamefont {Poon}, \citenamefont {Egelhaaf},\ and\ \citenamefont
  {Schurtenberger}}]{stradner04}%
  \BibitemOpen
  \bibfield  {author} {\bibinfo {author} {\bibfnamefont {A.}~\bibnamefont
  {Stradner}}, \bibinfo {author} {\bibfnamefont {H.}~\bibnamefont {Sedgwick}},
  \bibinfo {author} {\bibfnamefont {F.}~\bibnamefont {Cardinaux}}, \bibinfo
  {author} {\bibfnamefont {W.~C.}\ \bibnamefont {Poon}}, \bibinfo {author}
  {\bibfnamefont {S.~U.}\ \bibnamefont {Egelhaaf}}, \ and\ \bibinfo {author}
  {\bibfnamefont {P.}~\bibnamefont {Schurtenberger}},\ }\href@noop {}
  {\bibfield  {journal} {\bibinfo  {journal} {Nature}\ }\textbf {\bibinfo
  {volume} {432}},\ \bibinfo {pages} {492} (\bibinfo {year}
  {2004})}\BibitemShut {NoStop}%
\bibitem [{\citenamefont {Godfrin}\ \emph {et~al.}(2015)\citenamefont
  {Godfrin}, \citenamefont {Hudson}, \citenamefont {Hong}, \citenamefont
  {Porcar}, \citenamefont {Falus}, \citenamefont {Wagner},\ and\ \citenamefont
  {Liu}}]{godfrin2015}%
  \BibitemOpen
  \bibfield  {author} {\bibinfo {author} {\bibfnamefont {P.~D.}\ \bibnamefont
  {Godfrin}}, \bibinfo {author} {\bibfnamefont {S.~D.}\ \bibnamefont {Hudson}},
  \bibinfo {author} {\bibfnamefont {K.}~\bibnamefont {Hong}}, \bibinfo {author}
  {\bibfnamefont {L.}~\bibnamefont {Porcar}}, \bibinfo {author} {\bibfnamefont
  {P.}~\bibnamefont {Falus}}, \bibinfo {author} {\bibfnamefont {N.~J.}\
  \bibnamefont {Wagner}}, \ and\ \bibinfo {author} {\bibfnamefont
  {Y.}~\bibnamefont {Liu}},\ }\href@noop {} {\bibfield  {journal} {\bibinfo
  {journal} {Phys. Rev. Lett.}\ }\textbf {\bibinfo {volume} {115}},\ \bibinfo
  {pages} {228302} (\bibinfo {year} {2015})}\BibitemShut {NoStop}%
\bibitem [{\citenamefont {Schmidt}\ \emph {et~al.}(2008)\citenamefont
  {Schmidt}, \citenamefont {Guigas},\ and\ \citenamefont {Weiss}}]{schmidt08}%
  \BibitemOpen
  \bibfield  {author} {\bibinfo {author} {\bibfnamefont {U.}~\bibnamefont
  {Schmidt}}, \bibinfo {author} {\bibfnamefont {G.}~\bibnamefont {Guigas}}, \
  and\ \bibinfo {author} {\bibfnamefont {M.}~\bibnamefont {Weiss}},\
  }\href@noop {} {\bibfield  {journal} {\bibinfo  {journal} {Phys. Rev. Lett.}\
  }\textbf {\bibinfo {volume} {101}},\ \bibinfo {pages} {128104} (\bibinfo
  {year} {2008})}\BibitemShut {NoStop}%
\bibitem [{\citenamefont {Yan}\ \emph {et~al.}(2013)\citenamefont {Yan},
  \citenamefont {Enge}, \citenamefont {Whitington}, \citenamefont {Dave},
  \citenamefont {Liu}, \citenamefont {Sur}, \citenamefont {Schmierer},
  \citenamefont {Jolma}, \citenamefont {Kivioja},\ and\ \citenamefont
  {Taipale}}]{yan13}%
  \BibitemOpen
  \bibfield  {author} {\bibinfo {author} {\bibfnamefont {J.}~\bibnamefont
  {Yan}}, \bibinfo {author} {\bibfnamefont {M.}~\bibnamefont {Enge}}, \bibinfo
  {author} {\bibfnamefont {T.}~\bibnamefont {Whitington}}, \bibinfo {author}
  {\bibfnamefont {K.}~\bibnamefont {Dave}}, \bibinfo {author} {\bibfnamefont
  {J.}~\bibnamefont {Liu}}, \bibinfo {author} {\bibfnamefont {I.}~\bibnamefont
  {Sur}}, \bibinfo {author} {\bibfnamefont {B.}~\bibnamefont {Schmierer}},
  \bibinfo {author} {\bibfnamefont {A.}~\bibnamefont {Jolma}}, \bibinfo
  {author} {\bibfnamefont {T.}~\bibnamefont {Kivioja}}, \ and\ \bibinfo
  {author} {\bibfnamefont {M.}~\bibnamefont {Taipale}},\ }\href@noop {}
  {\bibfield  {journal} {\bibinfo  {journal} {Cell}\ }\textbf {\bibinfo
  {volume} {154}},\ \bibinfo {pages} {801} (\bibinfo {year}
  {2013})}\BibitemShut {NoStop}%
\bibitem [{\citenamefont {Monks}\ \emph {et~al.}(1998)\citenamefont {Monks},
  \citenamefont {Freiberg}, \citenamefont {Kupfer}, \citenamefont {Sciaky},\
  and\ \citenamefont {Kupfer}}]{monks98}%
  \BibitemOpen
  \bibfield  {author} {\bibinfo {author} {\bibfnamefont {C.~R.}\ \bibnamefont
  {Monks}}, \bibinfo {author} {\bibfnamefont {B.~A.}\ \bibnamefont {Freiberg}},
  \bibinfo {author} {\bibfnamefont {H.}~\bibnamefont {Kupfer}}, \bibinfo
  {author} {\bibfnamefont {N.}~\bibnamefont {Sciaky}}, \ and\ \bibinfo {author}
  {\bibfnamefont {A.}~\bibnamefont {Kupfer}},\ }\href@noop {} {\bibfield
  {journal} {\bibinfo  {journal} {Nature}\ }\textbf {\bibinfo {volume} {395}},\
  \bibinfo {pages} {82} (\bibinfo {year} {1998})}\BibitemShut {NoStop}%
\bibitem [{\citenamefont {Kakio}\ \emph {et~al.}(2001)\citenamefont {Kakio},
  \citenamefont {Nishimoto}, \citenamefont {Yanagisawa}, \citenamefont
  {Kozutsumi},\ and\ \citenamefont {Matsuzaki}}]{kakio01}%
  \BibitemOpen
  \bibfield  {author} {\bibinfo {author} {\bibfnamefont {A.}~\bibnamefont
  {Kakio}}, \bibinfo {author} {\bibfnamefont {S.-i.}\ \bibnamefont
  {Nishimoto}}, \bibinfo {author} {\bibfnamefont {K.}~\bibnamefont
  {Yanagisawa}}, \bibinfo {author} {\bibfnamefont {Y.}~\bibnamefont
  {Kozutsumi}}, \ and\ \bibinfo {author} {\bibfnamefont {K.}~\bibnamefont
  {Matsuzaki}},\ }\href@noop {} {\bibfield  {journal} {\bibinfo  {journal} {J.
  Biol. Chem.}\ }\textbf {\bibinfo {volume} {276}},\ \bibinfo {pages} {24985}
  (\bibinfo {year} {2001})}\BibitemShut {NoStop}%
\bibitem [{\citenamefont {Aguzzi}\ and\ \citenamefont
  {O'Connor}(2010)}]{aguzzi10}%
  \BibitemOpen
  \bibfield  {author} {\bibinfo {author} {\bibfnamefont {A.}~\bibnamefont
  {Aguzzi}}\ and\ \bibinfo {author} {\bibfnamefont {T.}~\bibnamefont
  {O'Connor}},\ }\href@noop {} {\bibfield  {journal} {\bibinfo  {journal} {Nat.
  Rev. Drug Discov}\ }\textbf {\bibinfo {volume} {9}},\ \bibinfo {pages} {237}
  (\bibinfo {year} {2010})}\BibitemShut {NoStop}%
\bibitem [{\citenamefont {Sear}(1999)}]{Sear1999}%
  \BibitemOpen
  \bibfield  {author} {\bibinfo {author} {\bibfnamefont {R.~P.}\ \bibnamefont
  {Sear}},\ }\href@noop {} {\bibfield  {journal} {\bibinfo  {journal} {J. Chem.
  Phys.}\ }\textbf {\bibinfo {volume} {111}},\ \bibinfo {pages} {4800}
  (\bibinfo {year} {1999})}\BibitemShut {NoStop}%
\bibitem [{\citenamefont {Goegelein}(2008)}]{Goegelein2008}%
  \BibitemOpen
  \bibfield  {author} {\bibinfo {author} {\bibfnamefont {C.}~\bibnamefont
  {Goegelein}},\ }\href@noop {} {\bibfield  {journal} {\bibinfo  {journal} {J.
  Chem. Phys.}\ }\textbf {\bibinfo {volume} {129}},\ \bibinfo {pages} {085102}
  (\bibinfo {year} {2008})}\BibitemShut {NoStop}%
\bibitem [{\citenamefont {Vagenende}\ \emph {et~al.}(2009)\citenamefont
  {Vagenende}, \citenamefont {Yap},\ and\ \citenamefont {Trout}}]{cosolvent}%
  \BibitemOpen
  \bibfield  {author} {\bibinfo {author} {\bibfnamefont {V.}~\bibnamefont
  {Vagenende}}, \bibinfo {author} {\bibfnamefont {M.~G.~S.}\ \bibnamefont
  {Yap}}, \ and\ \bibinfo {author} {\bibfnamefont {B.~L.}\ \bibnamefont
  {Trout}},\ }\href {\doibase 10.1021/bi900649t} {\bibfield  {journal}
  {\bibinfo  {journal} {Biochemistry}\ }\textbf {\bibinfo {volume} {48}},\
  \bibinfo {pages} {11084} (\bibinfo {year} {2009})}\BibitemShut {NoStop}%
\bibitem [{\citenamefont {Saricay}\ \emph {et~al.}(2012)\citenamefont
  {Saricay}, \citenamefont {Dhayal}, \citenamefont {Wierenga},\ and\
  \citenamefont {de~Vries}}]{saricay12}%
  \BibitemOpen
  \bibfield  {author} {\bibinfo {author} {\bibfnamefont {Y.}~\bibnamefont
  {Saricay}}, \bibinfo {author} {\bibfnamefont {S.~K.}\ \bibnamefont {Dhayal}},
  \bibinfo {author} {\bibfnamefont {P.~A.}\ \bibnamefont {Wierenga}}, \ and\
  \bibinfo {author} {\bibfnamefont {R.}~\bibnamefont {de~Vries}},\ }\href@noop
  {} {\bibfield  {journal} {\bibinfo  {journal} {Farad. Discuss}\ }\textbf
  {\bibinfo {volume} {158}},\ \bibinfo {pages} {51} (\bibinfo {year}
  {2012})}\BibitemShut {NoStop}%
\bibitem [{\citenamefont {Zhang}\ \emph {et~al.}(2008)\citenamefont {Zhang},
  \citenamefont {Skoda}, \citenamefont {Jacobs}, \citenamefont {Zorn},
  \citenamefont {Martin}, \citenamefont {Martin}, \citenamefont {Clark},
  \citenamefont {Weggler}, \citenamefont {Hildebrandt}, \citenamefont
  {Kohlbacher},\ and\ \citenamefont {Schreiber}}]{Zhang08}%
  \BibitemOpen
  \bibfield  {author} {\bibinfo {author} {\bibfnamefont {F.}~\bibnamefont
  {Zhang}}, \bibinfo {author} {\bibfnamefont {M.~W.~A.}\ \bibnamefont {Skoda}},
  \bibinfo {author} {\bibfnamefont {R.~M.~J.}\ \bibnamefont {Jacobs}}, \bibinfo
  {author} {\bibfnamefont {S.}~\bibnamefont {Zorn}}, \bibinfo {author}
  {\bibfnamefont {R.~A.}\ \bibnamefont {Martin}}, \bibinfo {author}
  {\bibfnamefont {C.~M.}\ \bibnamefont {Martin}}, \bibinfo {author}
  {\bibfnamefont {G.~F.}\ \bibnamefont {Clark}}, \bibinfo {author}
  {\bibfnamefont {S.}~\bibnamefont {Weggler}}, \bibinfo {author} {\bibfnamefont
  {A.}~\bibnamefont {Hildebrandt}}, \bibinfo {author} {\bibfnamefont
  {O.}~\bibnamefont {Kohlbacher}}, \ and\ \bibinfo {author} {\bibfnamefont
  {F.}~\bibnamefont {Schreiber}},\ }\href
  {http://link.aps.org/doi/10.1103/PhysRevLett.101.148101} {\bibfield
  {journal} {\bibinfo  {journal} {Phys. Rev. Lett.}\ }\textbf {\bibinfo
  {volume} {101}},\ \bibinfo {pages} {148101} (\bibinfo {year}
  {2008})}\BibitemShut {NoStop}%
\bibitem [{\citenamefont {Zhang}\ \emph {et~al.}(2012)\citenamefont {Zhang},
  \citenamefont {Roth}, \citenamefont {Wolf}, \citenamefont {Roosen-Runge},
  \citenamefont {Skoda}, \citenamefont {Jacobs}, \citenamefont {Stzucki},\ and\
  \citenamefont {Schreiber}}]{Zhang12}%
  \BibitemOpen
  \bibfield  {author} {\bibinfo {author} {\bibfnamefont {F.}~\bibnamefont
  {Zhang}}, \bibinfo {author} {\bibfnamefont {R.}~\bibnamefont {Roth}},
  \bibinfo {author} {\bibfnamefont {M.}~\bibnamefont {Wolf}}, \bibinfo {author}
  {\bibfnamefont {F.}~\bibnamefont {Roosen-Runge}}, \bibinfo {author}
  {\bibfnamefont {M.~W.~A.}\ \bibnamefont {Skoda}}, \bibinfo {author}
  {\bibfnamefont {R.~M.~J.}\ \bibnamefont {Jacobs}}, \bibinfo {author}
  {\bibfnamefont {M.}~\bibnamefont {Stzucki}}, \ and\ \bibinfo {author}
  {\bibfnamefont {F.}~\bibnamefont {Schreiber}},\ }\href {\doibase
  10.1039/C2SM07008A} {\bibfield  {journal} {\bibinfo  {journal} {Soft Matter}\
  }\textbf {\bibinfo {volume} {8}},\ \bibinfo {pages} {1313} (\bibinfo {year}
  {2012})}\BibitemShut {NoStop}%
\bibitem [{\citenamefont {Castner}(2017)}]{Castner17}%
  \BibitemOpen
  \bibfield  {author} {\bibinfo {author} {\bibfnamefont {D.~G.}\ \bibnamefont
  {Castner}},\ }\href@noop {} {\bibfield  {journal} {\bibinfo  {journal}
  {Biointerphases}\ }\textbf {\bibinfo {volume} {12}},\ \bibinfo {pages}
  {02C301} (\bibinfo {year} {2017})}\BibitemShut {NoStop}%
\bibitem [{\citenamefont {Sch{\"o}ttler}\ \emph {et~al.}(2016)\citenamefont
  {Sch{\"o}ttler}, \citenamefont {Becker}, \citenamefont {Winzen},
  \citenamefont {Steinbach}, \citenamefont {Mohr}, \citenamefont {Landfester},
  \citenamefont {Mail{\"a}nder},\ and\ \citenamefont {Wurm}}]{schottler16}%
  \BibitemOpen
  \bibfield  {author} {\bibinfo {author} {\bibfnamefont {S.}~\bibnamefont
  {Sch{\"o}ttler}}, \bibinfo {author} {\bibfnamefont {G.}~\bibnamefont
  {Becker}}, \bibinfo {author} {\bibfnamefont {S.}~\bibnamefont {Winzen}},
  \bibinfo {author} {\bibfnamefont {T.}~\bibnamefont {Steinbach}}, \bibinfo
  {author} {\bibfnamefont {K.}~\bibnamefont {Mohr}}, \bibinfo {author}
  {\bibfnamefont {K.}~\bibnamefont {Landfester}}, \bibinfo {author}
  {\bibfnamefont {V.}~\bibnamefont {Mail{\"a}nder}}, \ and\ \bibinfo {author}
  {\bibfnamefont {F.~R.}\ \bibnamefont {Wurm}},\ }\href@noop {} {\bibfield
  {journal} {\bibinfo  {journal} {Nat. Nanotech}\ } (\bibinfo {year}
  {2016})}\BibitemShut {NoStop}%
\bibitem [{\citenamefont {Su}\ \emph {et~al.}(1999)\citenamefont {Su},
  \citenamefont {Lu}, \citenamefont {Thomas},\ and\ \citenamefont
  {Cui}}]{Su99}%
  \BibitemOpen
  \bibfield  {author} {\bibinfo {author} {\bibfnamefont {T.~J.}\ \bibnamefont
  {Su}}, \bibinfo {author} {\bibfnamefont {J.~R.}\ \bibnamefont {Lu}}, \bibinfo
  {author} {\bibfnamefont {R.~K.}\ \bibnamefont {Thomas}}, \ and\ \bibinfo
  {author} {\bibfnamefont {Z.~F.}\ \bibnamefont {Cui}},\ }\href {\doibase
  10.1021/jp983580j} {\bibfield  {journal} {\bibinfo  {journal} {J. Phys. Chem.
  B}\ }\textbf {\bibinfo {volume} {103}},\ \bibinfo {pages} {3727} (\bibinfo
  {year} {1999})}\BibitemShut {NoStop}%
\bibitem [{\citenamefont {Lu}\ \emph {et~al.}(2007)\citenamefont {Lu},
  \citenamefont {Zhao},\ and\ \citenamefont {Yaseen}}]{Lu07}%
  \BibitemOpen
  \bibfield  {author} {\bibinfo {author} {\bibfnamefont {J.~R.}\ \bibnamefont
  {Lu}}, \bibinfo {author} {\bibfnamefont {X.}~\bibnamefont {Zhao}}, \ and\
  \bibinfo {author} {\bibfnamefont {M.}~\bibnamefont {Yaseen}},\ }\href
  {\doibase http://dx.doi.org/10.1016/j.cocis.2007.02.001} {\bibfield
  {journal} {\bibinfo  {journal} {Curr. Opin. Colloid Interface Sci.}\ }\textbf
  {\bibinfo {volume} {12}},\ \bibinfo {pages} {9} (\bibinfo {year}
  {2007})}\BibitemShut {NoStop}%
\bibitem [{\citenamefont {Gray}(2004)}]{Gray04}%
  \BibitemOpen
  \bibfield  {author} {\bibinfo {author} {\bibfnamefont {J.~J.}\ \bibnamefont
  {Gray}},\ }\href@noop {} {\bibfield  {journal} {\bibinfo  {journal} {Curr.
  Opin. Struct. Biol}\ }\textbf {\bibinfo {volume} {14}},\ \bibinfo {pages}
  {110} (\bibinfo {year} {2004})}\BibitemShut {NoStop}%
\bibitem [{SM()}]{SM}%
  \BibitemOpen
  \href@noop {} {\bibinfo  {journal} {Supporting Material}\ }\BibitemShut
  {NoStop}%
\bibitem [{\citenamefont {Kandori}\ \emph {et~al.}(2010)\citenamefont
  {Kandori}, \citenamefont {Toshima}, \citenamefont {Wakamura}, \citenamefont
  {Fukusumi},\ and\ \citenamefont {Morisada}}]{Kandori10}%
  \BibitemOpen
\bibfield  {journal} {  }\bibfield  {author} {\bibinfo {author} {\bibfnamefont
  {K.}~\bibnamefont {Kandori}}, \bibinfo {author} {\bibfnamefont
  {S.}~\bibnamefont {Toshima}}, \bibinfo {author} {\bibfnamefont
  {M.}~\bibnamefont {Wakamura}}, \bibinfo {author} {\bibfnamefont
  {M.}~\bibnamefont {Fukusumi}}, \ and\ \bibinfo {author} {\bibfnamefont
  {Y.}~\bibnamefont {Morisada}},\ }\href@noop {} {\bibfield  {journal}
  {\bibinfo  {journal} {J. Phys. Chem. B}\ }\textbf {\bibinfo {volume} {114}},\
  \bibinfo {pages} {2399} (\bibinfo {year} {2010})}\BibitemShut {NoStop}%
\bibitem [{\citenamefont {Anbazhagan}\ \emph {et~al.}(2016)\citenamefont
  {Anbazhagan}, \citenamefont {Rajendran}, \citenamefont {Natarajan},
  \citenamefont {Kiran},\ and\ \citenamefont {Pattanayak}}]{Anbazhagan16}%
  \BibitemOpen
  \bibfield  {author} {\bibinfo {author} {\bibfnamefont {E.}~\bibnamefont
  {Anbazhagan}}, \bibinfo {author} {\bibfnamefont {A.}~\bibnamefont
  {Rajendran}}, \bibinfo {author} {\bibfnamefont {D.}~\bibnamefont
  {Natarajan}}, \bibinfo {author} {\bibfnamefont {M.}~\bibnamefont {Kiran}}, \
  and\ \bibinfo {author} {\bibfnamefont {D.~K.}\ \bibnamefont {Pattanayak}},\
  }\href@noop {} {\bibfield  {journal} {\bibinfo  {journal} {Colloids Surf.,
  B}\ }\textbf {\bibinfo {volume} {143}},\ \bibinfo {pages} {213} (\bibinfo
  {year} {2016})}\BibitemShut {NoStop}%
\bibitem [{\citenamefont {Lee}\ \emph {et~al.}(2013)\citenamefont {Lee},
  \citenamefont {Schmidt}, \citenamefont {Laanait}, \citenamefont {Sturchio},\
  and\ \citenamefont {Fenter}}]{LeeSchmidtFenter2013}%
  \BibitemOpen
  \bibfield  {author} {\bibinfo {author} {\bibfnamefont {S.~S.}\ \bibnamefont
  {Lee}}, \bibinfo {author} {\bibfnamefont {M.}~\bibnamefont {Schmidt}},
  \bibinfo {author} {\bibfnamefont {N.}~\bibnamefont {Laanait}}, \bibinfo
  {author} {\bibfnamefont {N.~C.}\ \bibnamefont {Sturchio}}, \ and\ \bibinfo
  {author} {\bibfnamefont {P.}~\bibnamefont {Fenter}},\ }\href@noop {}
  {\bibfield  {journal} {\bibinfo  {journal} {J. Phys. Chem. C}\ }\textbf
  {\bibinfo {volume} {117}},\ \bibinfo {pages} {23738} (\bibinfo {year}
  {2013})}\BibitemShut {NoStop}%
\bibitem [{\citenamefont {Sauter}\ \emph {et~al.}(2015)\citenamefont {Sauter},
  \citenamefont {Roosen-Runge}, \citenamefont {Zhang}, \citenamefont {Lotze},
  \citenamefont {Jacobs},\ and\ \citenamefont {Schreiber}}]{Sauter15}%
  \BibitemOpen
  \bibfield  {author} {\bibinfo {author} {\bibfnamefont {A.}~\bibnamefont
  {Sauter}}, \bibinfo {author} {\bibfnamefont {F.}~\bibnamefont
  {Roosen-Runge}}, \bibinfo {author} {\bibfnamefont {F.}~\bibnamefont {Zhang}},
  \bibinfo {author} {\bibfnamefont {G.}~\bibnamefont {Lotze}}, \bibinfo
  {author} {\bibfnamefont {R.~M.~J.}\ \bibnamefont {Jacobs}}, \ and\ \bibinfo
  {author} {\bibfnamefont {F.}~\bibnamefont {Schreiber}},\ }\href@noop {}
  {\bibfield  {journal} {\bibinfo  {journal} {J. Am. Chem. Soc}\ }\textbf
  {\bibinfo {volume} {137}},\ \bibinfo {pages} {1485} (\bibinfo {year}
  {2015})}\BibitemShut {NoStop}%
\bibitem [{\citenamefont {Zhang}\ \emph {et~al.}(2010)\citenamefont {Zhang},
  \citenamefont {Weggler}, \citenamefont {Ziller}, \citenamefont {Ianeselli},
  \citenamefont {Heck}, \citenamefont {Hildebrandt}, \citenamefont
  {Kohlbacher}, \citenamefont {Skoda}, \citenamefont {Jacobs},\ and\
  \citenamefont {Schreiber}}]{zhang10}%
  \BibitemOpen
  \bibfield  {author} {\bibinfo {author} {\bibfnamefont {F.}~\bibnamefont
  {Zhang}}, \bibinfo {author} {\bibfnamefont {S.}~\bibnamefont {Weggler}},
  \bibinfo {author} {\bibfnamefont {M.~J.}\ \bibnamefont {Ziller}}, \bibinfo
  {author} {\bibfnamefont {L.}~\bibnamefont {Ianeselli}}, \bibinfo {author}
  {\bibfnamefont {B.~S.}\ \bibnamefont {Heck}}, \bibinfo {author}
  {\bibfnamefont {A.}~\bibnamefont {Hildebrandt}}, \bibinfo {author}
  {\bibfnamefont {O.}~\bibnamefont {Kohlbacher}}, \bibinfo {author}
  {\bibfnamefont {M.~W.}\ \bibnamefont {Skoda}}, \bibinfo {author}
  {\bibfnamefont {R.~M.}\ \bibnamefont {Jacobs}}, \ and\ \bibinfo {author}
  {\bibfnamefont {F.}~\bibnamefont {Schreiber}},\ }\href {\doibase
  10.1002/prot.22852} {\bibfield  {journal} {\bibinfo  {journal} {Proteins}\
  }\textbf {\bibinfo {volume} {78}},\ \bibinfo {pages} {3450} (\bibinfo {year}
  {2010})}\BibitemShut {NoStop}%
\bibitem [{\citenamefont {Roosen-Runge}\ \emph {et~al.}(2014)\citenamefont
  {Roosen-Runge}, \citenamefont {Zhang}, \citenamefont {Schreiber},\ and\
  \citenamefont {Roth}}]{RoosenRunge2014}%
  \BibitemOpen
  \bibfield  {author} {\bibinfo {author} {\bibfnamefont {F.}~\bibnamefont
  {Roosen-Runge}}, \bibinfo {author} {\bibfnamefont {F.}~\bibnamefont {Zhang}},
  \bibinfo {author} {\bibfnamefont {F.}~\bibnamefont {Schreiber}}, \ and\
  \bibinfo {author} {\bibfnamefont {R.}~\bibnamefont {Roth}},\ }\href@noop {}
  {\bibfield  {journal} {\bibinfo  {journal} {Sci. Rep.}\ }\textbf {\bibinfo
  {volume} {4}},\ \bibinfo {pages} {7016} (\bibinfo {year} {2014})}\BibitemShut
  {NoStop}%
\bibitem [{\citenamefont {Braun}\ \emph {et~al.}(2017)\citenamefont {Braun},
  \citenamefont {Wolf}, \citenamefont {Matsarskaia}, \citenamefont {Da~Vela},
  \citenamefont {Roosen-Runge}, \citenamefont {Sztucki}, \citenamefont {Roth},
  \citenamefont {Zhang},\ and\ \citenamefont {Schreiber}}]{Braun17}%
  \BibitemOpen
  \bibfield  {author} {\bibinfo {author} {\bibfnamefont {M.~K.}\ \bibnamefont
  {Braun}}, \bibinfo {author} {\bibfnamefont {M.}~\bibnamefont {Wolf}},
  \bibinfo {author} {\bibfnamefont {O.}~\bibnamefont {Matsarskaia}}, \bibinfo
  {author} {\bibfnamefont {S.}~\bibnamefont {Da~Vela}}, \bibinfo {author}
  {\bibfnamefont {F.}~\bibnamefont {Roosen-Runge}}, \bibinfo {author}
  {\bibfnamefont {M.}~\bibnamefont {Sztucki}}, \bibinfo {author} {\bibfnamefont
  {R.}~\bibnamefont {Roth}}, \bibinfo {author} {\bibfnamefont {F.}~\bibnamefont
  {Zhang}}, \ and\ \bibinfo {author} {\bibfnamefont {F.}~\bibnamefont
  {Schreiber}},\ }\href@noop {} {\bibfield  {journal} {\bibinfo  {journal} {J.
  Phys. Chem. B}\ }\textbf {\bibinfo {volume} {121}},\ \bibinfo {pages} {1731}
  (\bibinfo {year} {2017})}\BibitemShut {NoStop}%
\bibitem [{\citenamefont {Zhang}\ \emph {et~al.}(2014)\citenamefont {Zhang},
  \citenamefont {Roosen-Runge}, \citenamefont {Sauter}, \citenamefont {Wolf},
  \citenamefont {Jacobs},\ and\ \citenamefont {Schreiber}}]{Zhang14}%
  \BibitemOpen
  \bibfield  {author} {\bibinfo {author} {\bibfnamefont {F.}~\bibnamefont
  {Zhang}}, \bibinfo {author} {\bibfnamefont {F.}~\bibnamefont {Roosen-Runge}},
  \bibinfo {author} {\bibfnamefont {A.}~\bibnamefont {Sauter}}, \bibinfo
  {author} {\bibfnamefont {M.}~\bibnamefont {Wolf}}, \bibinfo {author}
  {\bibfnamefont {R.~M.~J.}\ \bibnamefont {Jacobs}}, \ and\ \bibinfo {author}
  {\bibfnamefont {F.}~\bibnamefont {Schreiber}},\ }\href {\doibase
  10.1515/pac-2014-5002} {\bibfield  {journal} {\bibinfo  {journal} {Pure Appl.
  Chem.}\ }\textbf {\bibinfo {volume} {86}},\ \bibinfo {pages} {191} (\bibinfo
  {year} {2014})}\BibitemShut {NoStop}%
\bibitem [{\citenamefont {Hlady}\ and\ \citenamefont
  {Andrade}(1988)}]{Hlady88}%
  \BibitemOpen
  \bibfield  {author} {\bibinfo {author} {\bibfnamefont {V.}~\bibnamefont
  {Hlady}}\ and\ \bibinfo {author} {\bibfnamefont {J.~D.}\ \bibnamefont
  {Andrade}},\ }\href {\doibase http://dx.doi.org/10.1016/0166-6622(88)80029-5}
  {\bibfield  {journal} {\bibinfo  {journal} {Colloids Surf.}\ }\textbf
  {\bibinfo {volume} {32}},\ \bibinfo {pages} {359} (\bibinfo {year}
  {1988})}\BibitemShut {NoStop}%
\bibitem [{\citenamefont {Tilton}\ \emph {et~al.}(1990)\citenamefont {Tilton},
  \citenamefont {Robertson},\ and\ \citenamefont {Gast}}]{Tilton90}%
  \BibitemOpen
  \bibfield  {author} {\bibinfo {author} {\bibfnamefont {R.~D.}\ \bibnamefont
  {Tilton}}, \bibinfo {author} {\bibfnamefont {C.~R.}\ \bibnamefont
  {Robertson}}, \ and\ \bibinfo {author} {\bibfnamefont {A.~P.}\ \bibnamefont
  {Gast}},\ }\href@noop {} {\bibfield  {journal} {\bibinfo  {journal} {J.
  Colloid Interface Sci.}\ }\textbf {\bibinfo {volume} {137}},\ \bibinfo
  {pages} {192} (\bibinfo {year} {1990})}\BibitemShut {NoStop}%
\bibitem [{\citenamefont {Rabe}\ \emph {et~al.}(2009)\citenamefont {Rabe},
  \citenamefont {Verdes},\ and\ \citenamefont {Seeger}}]{Rabe09}%
  \BibitemOpen
  \bibfield  {author} {\bibinfo {author} {\bibfnamefont {M.}~\bibnamefont
  {Rabe}}, \bibinfo {author} {\bibfnamefont {D.}~\bibnamefont {Verdes}}, \ and\
  \bibinfo {author} {\bibfnamefont {S.}~\bibnamefont {Seeger}},\ }\href@noop {}
  {\bibfield  {journal} {\bibinfo  {journal} {Soft Matter}\ }\textbf {\bibinfo
  {volume} {5}},\ \bibinfo {pages} {1039} (\bibinfo {year} {2009})}\BibitemShut
  {NoStop}%
\bibitem [{\citenamefont {Su}\ \emph {et~al.}(1998)\citenamefont {Su},
  \citenamefont {Lu}, \citenamefont {Thomas}, \citenamefont {Cui},\ and\
  \citenamefont {Penfold}}]{Su98}%
  \BibitemOpen
  \bibfield  {author} {\bibinfo {author} {\bibfnamefont {T.~J.}\ \bibnamefont
  {Su}}, \bibinfo {author} {\bibnamefont {Lu}}, \bibinfo {author}
  {\bibfnamefont {R.~K.}\ \bibnamefont {Thomas}}, \bibinfo {author}
  {\bibfnamefont {Z.~F.}\ \bibnamefont {Cui}}, \ and\ \bibinfo {author}
  {\bibfnamefont {J.}~\bibnamefont {Penfold}},\ }\href {\doibase
  10.1021/jp981239t} {\bibfield  {journal} {\bibinfo  {journal} {J. Phys. Chem.
  B}\ }\textbf {\bibinfo {volume} {102}},\ \bibinfo {pages} {8100} (\bibinfo
  {year} {1998})}\BibitemShut {NoStop}%
\bibitem [{\citenamefont {Rabe}\ \emph {et~al.}(2008)\citenamefont {Rabe},
  \citenamefont {Verdes}, \citenamefont {Zimmermann},\ and\ \citenamefont
  {Seeger}}]{Rabe08}%
  \BibitemOpen
  \bibfield  {author} {\bibinfo {author} {\bibfnamefont {M.}~\bibnamefont
  {Rabe}}, \bibinfo {author} {\bibfnamefont {D.}~\bibnamefont {Verdes}},
  \bibinfo {author} {\bibfnamefont {J.}~\bibnamefont {Zimmermann}}, \ and\
  \bibinfo {author} {\bibfnamefont {S.}~\bibnamefont {Seeger}},\ }\href
  {\doibase 10.1021/jp804532v} {\bibfield  {journal} {\bibinfo  {journal} {J.
  Phys. Chem. B}\ }\textbf {\bibinfo {volume} {112}},\ \bibinfo {pages} {13971}
  (\bibinfo {year} {2008})}\BibitemShut {NoStop}%
\bibitem [{\citenamefont {Larsericsdotter}\ \emph {et~al.}(2005)\citenamefont
  {Larsericsdotter}, \citenamefont {Oscarsson},\ and\ \citenamefont
  {Buijs}}]{Larsericsdotter05}%
  \BibitemOpen
  \bibfield  {author} {\bibinfo {author} {\bibfnamefont {H.}~\bibnamefont
  {Larsericsdotter}}, \bibinfo {author} {\bibfnamefont {S.}~\bibnamefont
  {Oscarsson}}, \ and\ \bibinfo {author} {\bibfnamefont {J.}~\bibnamefont
  {Buijs}},\ }\href@noop {} {\bibfield  {journal} {\bibinfo  {journal} {J.
  Colloid Interface Sci.}\ }\textbf {\bibinfo {volume} {289}},\ \bibinfo
  {pages} {26} (\bibinfo {year} {2005})}\BibitemShut {NoStop}%
\bibitem [{\citenamefont {Giacomelli}\ and\ \citenamefont
  {Norde}(2001)}]{Giacomelli01}%
  \BibitemOpen
  \bibfield  {author} {\bibinfo {author} {\bibfnamefont {C.~E.}\ \bibnamefont
  {Giacomelli}}\ and\ \bibinfo {author} {\bibfnamefont {W.}~\bibnamefont
  {Norde}},\ }\href {\doibase http://dx.doi.org/10.1006/jcis.2000.7219}
  {\bibfield  {journal} {\bibinfo  {journal} {J. Colloid Interface Sci}\
  }\textbf {\bibinfo {volume} {233}},\ \bibinfo {pages} {234} (\bibinfo {year}
  {2001})}\BibitemShut {NoStop}%
\bibitem [{\citenamefont {Tarasevich}\ and\ \citenamefont
  {Monakhova}(2002)}]{Tarasevich02}%
  \BibitemOpen
  \bibfield  {author} {\bibinfo {author} {\bibfnamefont {Y.~I.}\ \bibnamefont
  {Tarasevich}}\ and\ \bibinfo {author} {\bibfnamefont {L.~I.}\ \bibnamefont
  {Monakhova}},\ }\href {\doibase 10.1023/a:1016828322434} {\bibfield
  {journal} {\bibinfo  {journal} {Colloid J.}\ }\textbf {\bibinfo {volume}
  {64}},\ \bibinfo {pages} {482} (\bibinfo {year} {2002})}\BibitemShut
  {NoStop}%
\bibitem [{\citenamefont {Roach}\ \emph {et~al.}(2005)\citenamefont {Roach},
  \citenamefont {Farrar},\ and\ \citenamefont {Perry}}]{Roach05}%
  \BibitemOpen
  \bibfield  {author} {\bibinfo {author} {\bibfnamefont {P.}~\bibnamefont
  {Roach}}, \bibinfo {author} {\bibfnamefont {D.}~\bibnamefont {Farrar}}, \
  and\ \bibinfo {author} {\bibfnamefont {C.~C.}\ \bibnamefont {Perry}},\ }\href
  {\doibase 10.1021/ja042898o} {\bibfield  {journal} {\bibinfo  {journal} {J.
  Am. Chem. Soc}\ }\textbf {\bibinfo {volume} {127}},\ \bibinfo {pages} {8168}
  (\bibinfo {year} {2005})}\BibitemShut {NoStop}%
\bibitem [{\citenamefont {Mondon}\ \emph {et~al.}(2003)\citenamefont {Mondon},
  \citenamefont {Berger},\ and\ \citenamefont {Ziegler}}]{mondon03}%
  \BibitemOpen
  \bibfield  {author} {\bibinfo {author} {\bibfnamefont {M.}~\bibnamefont
  {Mondon}}, \bibinfo {author} {\bibfnamefont {S.}~\bibnamefont {Berger}}, \
  and\ \bibinfo {author} {\bibfnamefont {C.}~\bibnamefont {Ziegler}},\
  }\href@noop {} {\bibfield  {journal} {\bibinfo  {journal} {Anal. Bioanal.
  Chem}\ }\textbf {\bibinfo {volume} {375}},\ \bibinfo {pages} {849} (\bibinfo
  {year} {2003})}\BibitemShut {NoStop}%
\bibitem [{\citenamefont {Elwing}(1998)}]{Elwing98}%
  \BibitemOpen
  \bibfield  {author} {\bibinfo {author} {\bibfnamefont {H.}~\bibnamefont
  {Elwing}},\ }\href@noop {} {\bibfield  {journal} {\bibinfo  {journal}
  {Biomaterials}\ }\textbf {\bibinfo {volume} {19}},\ \bibinfo {pages} {397}
  (\bibinfo {year} {1998})}\BibitemShut {NoStop}%
\bibitem [{\citenamefont {Tsargorodskaya}\ \emph {et~al.}(2004)\citenamefont
  {Tsargorodskaya}, \citenamefont {Nabok},\ and\ \citenamefont
  {Ray}}]{Tsargorodskaya04}%
  \BibitemOpen
  \bibfield  {author} {\bibinfo {author} {\bibfnamefont {A.}~\bibnamefont
  {Tsargorodskaya}}, \bibinfo {author} {\bibfnamefont {A.~V.}\ \bibnamefont
  {Nabok}}, \ and\ \bibinfo {author} {\bibfnamefont {A.~K.}\ \bibnamefont
  {Ray}},\ }\href {http://stacks.iop.org/0957-4484/15/i=5/a=051} {\bibfield
  {journal} {\bibinfo  {journal} {Nanotechnology}\ }\textbf {\bibinfo {volume}
  {15}},\ \bibinfo {pages} {703} (\bibinfo {year} {2004})}\BibitemShut
  {NoStop}%
\bibitem [{\citenamefont {Sauter}(2011)}]{Sauter11}%
  \BibitemOpen
  \bibfield  {author} {\bibinfo {author} {\bibfnamefont {A.}~\bibnamefont
  {Sauter}},\ }\emph {\bibinfo {title} {Crystallization and phase behavior of
  aqueous beta-Lactoglobulin solutions in the presence of multivalent
  cations}},\ \href@noop {} {\bibinfo {type} {Diploma thesis}} (\bibinfo {year}
  {2011})\BibitemShut {NoStop}%
\bibitem [{\citenamefont {Herzinger}\ \emph {et~al.}(1998)\citenamefont
  {Herzinger}, \citenamefont {Johs}, \citenamefont {McGahan}, \citenamefont
  {Woollam},\ and\ \citenamefont {Paulson}}]{herzinger98}%
  \BibitemOpen
  \bibfield  {author} {\bibinfo {author} {\bibfnamefont {C.}~\bibnamefont
  {Herzinger}}, \bibinfo {author} {\bibfnamefont {B.}~\bibnamefont {Johs}},
  \bibinfo {author} {\bibfnamefont {W.}~\bibnamefont {McGahan}}, \bibinfo
  {author} {\bibfnamefont {J.~A.}\ \bibnamefont {Woollam}}, \ and\ \bibinfo
  {author} {\bibfnamefont {W.}~\bibnamefont {Paulson}},\ }\href@noop {}
  {\bibfield  {journal} {\bibinfo  {journal} {J. Appl. Phys}\ }\textbf
  {\bibinfo {volume} {83}},\ \bibinfo {pages} {3323} (\bibinfo {year}
  {1998})}\BibitemShut {NoStop}%
\bibitem [{\citenamefont {Schiebener}\ \emph {et~al.}(1990)\citenamefont
  {Schiebener}, \citenamefont {Straub}, \citenamefont {Levelt~Sengers},\ and\
  \citenamefont {Gallagher}}]{schiebener90}%
  \BibitemOpen
  \bibfield  {author} {\bibinfo {author} {\bibfnamefont {P.}~\bibnamefont
  {Schiebener}}, \bibinfo {author} {\bibfnamefont {J.}~\bibnamefont {Straub}},
  \bibinfo {author} {\bibfnamefont {J.}~\bibnamefont {Levelt~Sengers}}, \ and\
  \bibinfo {author} {\bibfnamefont {J.}~\bibnamefont {Gallagher}},\ }\href@noop
  {} {\bibfield  {journal} {\bibinfo  {journal} {J. Phys. Chem. Ref. Data}\
  }\textbf {\bibinfo {volume} {19}},\ \bibinfo {pages} {677} (\bibinfo {year}
  {1990})}\BibitemShut {NoStop}%
\bibitem [{\citenamefont {Voinova}\ \emph {et~al.}(2002)\citenamefont
  {Voinova}, \citenamefont {Jonson},\ and\ \citenamefont {Kasemo}}]{voinova02}%
  \BibitemOpen
  \bibfield  {author} {\bibinfo {author} {\bibfnamefont {M.}~\bibnamefont
  {Voinova}}, \bibinfo {author} {\bibfnamefont {M.}~\bibnamefont {Jonson}}, \
  and\ \bibinfo {author} {\bibfnamefont {B.}~\bibnamefont {Kasemo}},\
  }\href@noop {} {\bibfield  {journal} {\bibinfo  {journal} {Biosens.
  Bioelectron}\ }\textbf {\bibinfo {volume} {17}},\ \bibinfo {pages} {835}
  (\bibinfo {year} {2002})}\BibitemShut {NoStop}%
\bibitem [{\citenamefont {Feiler}\ \emph {et~al.}(2007)\citenamefont {Feiler},
  \citenamefont {Sahlholm}, \citenamefont {Sandberg},\ and\ \citenamefont
  {Caldwell}}]{feiler07}%
  \BibitemOpen
  \bibfield  {author} {\bibinfo {author} {\bibfnamefont {A.~A.}\ \bibnamefont
  {Feiler}}, \bibinfo {author} {\bibfnamefont {A.}~\bibnamefont {Sahlholm}},
  \bibinfo {author} {\bibfnamefont {T.}~\bibnamefont {Sandberg}}, \ and\
  \bibinfo {author} {\bibfnamefont {K.~D.}\ \bibnamefont {Caldwell}},\
  }\href@noop {} {\bibfield  {journal} {\bibinfo  {journal} {J. Colloid
  Interface Sci}\ }\textbf {\bibinfo {volume} {315}},\ \bibinfo {pages} {475}
  (\bibinfo {year} {2007})}\BibitemShut {NoStop}%
\bibitem [{\citenamefont {Webster}\ \emph {et~al.}(2006)\citenamefont
  {Webster}, \citenamefont {Holt},\ and\ \citenamefont
  {Dalgliesh}}]{Webster06}%
  \BibitemOpen
  \bibfield  {author} {\bibinfo {author} {\bibfnamefont {J.}~\bibnamefont
  {Webster}}, \bibinfo {author} {\bibfnamefont {S.}~\bibnamefont {Holt}}, \
  and\ \bibinfo {author} {\bibfnamefont {R.}~\bibnamefont {Dalgliesh}},\ }\href
  {\doibase http://dx.doi.org/10.1016/j.physb.2006.05.400} {\bibfield
  {journal} {\bibinfo  {journal} {Phys. B: Condens. Matter}\ ,\ \bibinfo
  {pages} {1164}} (\bibinfo {year} {2006})}\BibitemShut {NoStop}%
\bibitem [{\citenamefont {Skoda}\ \emph {et~al.}(2009)\citenamefont {Skoda},
  \citenamefont {Schreiber}, \citenamefont {Jacobs}, \citenamefont {Webster},
  \citenamefont {Wolff}, \citenamefont {Dahint}, \citenamefont {Schwendel},\
  and\ \citenamefont {Grunze}}]{Skoda09}%
  \BibitemOpen
  \bibfield  {author} {\bibinfo {author} {\bibfnamefont {M.}~\bibnamefont
  {Skoda}}, \bibinfo {author} {\bibfnamefont {F.}~\bibnamefont {Schreiber}},
  \bibinfo {author} {\bibfnamefont {R.}~\bibnamefont {Jacobs}}, \bibinfo
  {author} {\bibfnamefont {J.}~\bibnamefont {Webster}}, \bibinfo {author}
  {\bibfnamefont {M.}~\bibnamefont {Wolff}}, \bibinfo {author} {\bibfnamefont
  {R.}~\bibnamefont {Dahint}}, \bibinfo {author} {\bibfnamefont
  {D.}~\bibnamefont {Schwendel}}, \ and\ \bibinfo {author} {\bibfnamefont
  {M.}~\bibnamefont {Grunze}},\ }\href@noop {} {\bibfield  {journal} {\bibinfo
  {journal} {Langmuir}\ }\textbf {\bibinfo {volume} {25}},\ \bibinfo {pages}
  {4056} (\bibinfo {year} {2009})}\BibitemShut {NoStop}%
\bibitem [{\citenamefont {Yohannes}\ \emph {et~al.}(2010)\citenamefont
  {Yohannes}, \citenamefont {Wiedmer}, \citenamefont {Elomaa}, \citenamefont
  {Jussila}, \citenamefont {Aseyev},\ and\ \citenamefont
  {Riekkola}}]{Yohannes10}%
  \BibitemOpen
  \bibfield  {author} {\bibinfo {author} {\bibfnamefont {G.}~\bibnamefont
  {Yohannes}}, \bibinfo {author} {\bibfnamefont {S.~K.}\ \bibnamefont
  {Wiedmer}}, \bibinfo {author} {\bibfnamefont {M.}~\bibnamefont {Elomaa}},
  \bibinfo {author} {\bibfnamefont {M.}~\bibnamefont {Jussila}}, \bibinfo
  {author} {\bibfnamefont {V.}~\bibnamefont {Aseyev}}, \ and\ \bibinfo {author}
  {\bibfnamefont {M.~L.}\ \bibnamefont {Riekkola}},\ }\href {\doibase
  10.1016/j.aca.2010.07.016} {\bibfield  {journal} {\bibinfo  {journal} {Ana.
  Chim. Acta}\ }\textbf {\bibinfo {volume} {675}},\ \bibinfo {pages} {191}
  (\bibinfo {year} {2010})}\BibitemShut {NoStop}%
\bibitem [{\citenamefont {Jackson}\ \emph {et~al.}(1988)\citenamefont
  {Jackson}, \citenamefont {Chapman},\ and\ \citenamefont
  {Gubbins}}]{Jackson1988}%
  \BibitemOpen
  \bibfield  {author} {\bibinfo {author} {\bibfnamefont {G.}~\bibnamefont
  {Jackson}}, \bibinfo {author} {\bibfnamefont {W.}~\bibnamefont {Chapman}}, \
  and\ \bibinfo {author} {\bibfnamefont {K.}~\bibnamefont {Gubbins}},\
  }\href@noop {} {\bibfield  {journal} {\bibinfo  {journal} {Mol. Phys.}\
  }\textbf {\bibinfo {volume} {65}},\ \bibinfo {pages} {1} (\bibinfo {year}
  {1988})}\BibitemShut {NoStop}%
\bibitem [{\citenamefont {Chapman}\ \emph {et~al.}(1988)\citenamefont
  {Chapman}, \citenamefont {Jackson},\ and\ \citenamefont
  {Gubbins}}]{Chapman1988}%
  \BibitemOpen
  \bibfield  {author} {\bibinfo {author} {\bibfnamefont {W.}~\bibnamefont
  {Chapman}}, \bibinfo {author} {\bibfnamefont {G.}~\bibnamefont {Jackson}}, \
  and\ \bibinfo {author} {\bibfnamefont {K.}~\bibnamefont {Gubbins}},\
  }\href@noop {} {\bibfield  {journal} {\bibinfo  {journal} {Mol. Phys.}\
  }\textbf {\bibinfo {volume} {65}},\ \bibinfo {pages} {1057} (\bibinfo {year}
  {1988})}\BibitemShut {NoStop}%
\bibitem [{\citenamefont {Wertheim}(1984{\natexlab{a}})}]{WertheimI1984}%
  \BibitemOpen
  \bibfield  {author} {\bibinfo {author} {\bibfnamefont {M.}~\bibnamefont
  {Wertheim}},\ }\href@noop {} {\bibfield  {journal} {\bibinfo  {journal} {J.
  Stat. Phys.}\ }\textbf {\bibinfo {volume} {35}},\ \bibinfo {pages} {19}
  (\bibinfo {year} {1984}{\natexlab{a}})}\BibitemShut {NoStop}%
\bibitem [{\citenamefont {Wertheim}(1984{\natexlab{b}})}]{WertheimII1984}%
  \BibitemOpen
  \bibfield  {author} {\bibinfo {author} {\bibfnamefont {M.}~\bibnamefont
  {Wertheim}},\ }\href@noop {} {\bibfield  {journal} {\bibinfo  {journal} {J.
  Stat. Phys.}\ }\textbf {\bibinfo {volume} {35}},\ \bibinfo {pages} {35}
  (\bibinfo {year} {1984}{\natexlab{b}})}\BibitemShut {NoStop}%
\bibitem [{\citenamefont {Wertheim}(1986{\natexlab{a}})}]{WertheimIII1986}%
  \BibitemOpen
  \bibfield  {author} {\bibinfo {author} {\bibfnamefont {M.}~\bibnamefont
  {Wertheim}},\ }\href@noop {} {\bibfield  {journal} {\bibinfo  {journal} {J.
  Stat. Phys.}\ }\textbf {\bibinfo {volume} {42}},\ \bibinfo {pages} {459}
  (\bibinfo {year} {1986}{\natexlab{a}})}\BibitemShut {NoStop}%
\bibitem [{\citenamefont {Wertheim}(1986{\natexlab{b}})}]{WertheimIV1986}%
  \BibitemOpen
  \bibfield  {author} {\bibinfo {author} {\bibfnamefont {M.}~\bibnamefont
  {Wertheim}},\ }\href@noop {} {\bibfield  {journal} {\bibinfo  {journal} {J.
  Stat. Phys.}\ }\textbf {\bibinfo {volume} {42}},\ \bibinfo {pages} {477}
  (\bibinfo {year} {1986}{\natexlab{b}})}\BibitemShut {NoStop}%
\bibitem [{\citenamefont {Romano}\ \emph {et~al.}(2010)\citenamefont {Romano},
  \citenamefont {Sanz},\ and\ \citenamefont {Sciortino}}]{romano2010phase}%
  \BibitemOpen
  \bibfield  {author} {\bibinfo {author} {\bibfnamefont {F.}~\bibnamefont
  {Romano}}, \bibinfo {author} {\bibfnamefont {E.}~\bibnamefont {Sanz}}, \ and\
  \bibinfo {author} {\bibfnamefont {F.}~\bibnamefont {Sciortino}},\ }\href@noop
  {} {\bibfield  {journal} {\bibinfo  {journal} {J. Chem. Phys.}\ }\textbf
  {\bibinfo {volume} {132}},\ \bibinfo {pages} {184501} (\bibinfo {year}
  {2010})}\BibitemShut {NoStop}%
\bibitem [{\citenamefont {Russo}\ \emph {et~al.}(2011)\citenamefont {Russo},
  \citenamefont {Tavares}, \citenamefont {Teixeira}, \citenamefont {Da~Gama},\
  and\ \citenamefont {Sciortino}}]{russo2011re}%
  \BibitemOpen
  \bibfield  {author} {\bibinfo {author} {\bibfnamefont {J.}~\bibnamefont
  {Russo}}, \bibinfo {author} {\bibfnamefont {J.}~\bibnamefont {Tavares}},
  \bibinfo {author} {\bibfnamefont {P.~I.~C.}\ \bibnamefont {Teixeira}},
  \bibinfo {author} {\bibfnamefont {M.~T.}\ \bibnamefont {Da~Gama}}, \ and\
  \bibinfo {author} {\bibfnamefont {F.}~\bibnamefont {Sciortino}},\ }\href@noop
  {} {\bibfield  {journal} {\bibinfo  {journal} {J. Chem. Phys.}\ }\textbf
  {\bibinfo {volume} {135}},\ \bibinfo {pages} {034501} (\bibinfo {year}
  {2011})}\BibitemShut {NoStop}%
\bibitem [{\citenamefont {Pawar}\ and\ \citenamefont
  {Kretzschmar}(2010)}]{pawar2010fabrication}%
  \BibitemOpen
  \bibfield  {author} {\bibinfo {author} {\bibfnamefont {A.~B.}\ \bibnamefont
  {Pawar}}\ and\ \bibinfo {author} {\bibfnamefont {I.}~\bibnamefont
  {Kretzschmar}},\ }\href@noop {} {\bibfield  {journal} {\bibinfo  {journal}
  {Macromol. Rapid Commun.}\ }\textbf {\bibinfo {volume} {31}},\ \bibinfo
  {pages} {150} (\bibinfo {year} {2010})}\BibitemShut {NoStop}%
\bibitem [{\citenamefont {Evans}(1979)}]{Evans1979}%
  \BibitemOpen
  \bibfield  {author} {\bibinfo {author} {\bibfnamefont {R.}~\bibnamefont
  {Evans}},\ }\href@noop {} {\bibfield  {journal} {\bibinfo  {journal} {Adv.
  Phys.}\ }\textbf {\bibinfo {volume} {28}},\ \bibinfo {pages} {143} (\bibinfo
  {year} {1979})}\BibitemShut {NoStop}%
\bibitem [{\citenamefont {Yu}\ and\ \citenamefont {Wu}(2002)}]{YuWu2002}%
  \BibitemOpen
  \bibfield  {author} {\bibinfo {author} {\bibfnamefont {Y.-X.}\ \bibnamefont
  {Yu}}\ and\ \bibinfo {author} {\bibfnamefont {J.}~\bibnamefont {Wu}},\
  }\href@noop {} {\bibfield  {journal} {\bibinfo  {journal} {J. Chem. Phys.}\
  }\textbf {\bibinfo {volume} {116}},\ \bibinfo {pages} {7094} (\bibinfo {year}
  {2002})}\BibitemShut {NoStop}%
\bibitem [{\citenamefont {Rosenfeld}(1989)}]{Rosenfeld1989}%
  \BibitemOpen
  \bibfield  {author} {\bibinfo {author} {\bibfnamefont {Y.}~\bibnamefont
  {Rosenfeld}},\ }\href@noop {} {\bibfield  {journal} {\bibinfo  {journal}
  {Phys. Rev. Lett.}\ }\textbf {\bibinfo {volume} {63}},\ \bibinfo {pages}
  {980} (\bibinfo {year} {1989})}\BibitemShut {NoStop}%
\bibitem [{\citenamefont {Roth}(2010)}]{Roth2010}%
  \BibitemOpen
  \bibfield  {author} {\bibinfo {author} {\bibfnamefont {R.}~\bibnamefont
  {Roth}},\ }\href@noop {} {\bibfield  {journal} {\bibinfo  {journal} {J.
  Phys.: Condens. Matter}\ }\textbf {\bibinfo {volume} {22}},\ \bibinfo {pages}
  {063102} (\bibinfo {year} {2010})}\BibitemShut {NoStop}%
\end{thebibliography}%

\end{document}